\documentstyle[amstex,prl,aps,epsfig]{revtex}

\begin{document}

\twocolumn

\noindent{\bf Rigorous Bounds to Retarded Learning}
 \\

Using an elegant approach, Herschkowitz and Opper~\cite{HO} 
established rigourous bounds on the information inferable
(learnable) from a set of data ($m$ points $\in {\Re}^N$) when
the latter are drawn from a distribution $P({\bf x})=P_0({\bf
x})\; \exp[-V(\lambda)]$, where $P_0({\bf x})$ is a spherical
normal law and $\exp[-V(\lambda)]$ is a modulation along an 
unknown anisotropy axis $\lambda={\bf w} \cdot {\bf x}$,
for some direction $\bf w$. They show in particular 
that if $P({\bf x})$ has zero mean, it is
impossible to learn the direction of anisotropy below a critical
fraction of data $\alpha^*$, and claim that 
$\alpha^*=\alpha_{lb} \equiv (1-\overline{\lambda^2})^{-1}$ {\it only} 
depends on $\overline{\lambda^2}$, the second moment of the distribution 
along $\lambda$, $P(\lambda) \equiv e^{-\lambda^2/2} \exp[-V(\lambda)]/\sqrt{2 \pi}$. 

The authors reach this conclusion by an expansion at small
$q$ of the upper bound to $\Delta R$, the difference between the
trivial risk and the cumulative Bayes risk. In the thermodynamic
limit ($m \rightarrow \infty$, $N \rightarrow \infty$ with
$\alpha=m/N$ finite) the upper bound$^[$\footnote{$\min_q$ erroneously 
stands for $\max_q$ in eq.(7) of~\cite{HO}}$^]$ 
is given by $\max_q G_{\alpha}(q)=G_{\alpha}(q^*)$, with
$G_{\alpha}(q) = \ln(1-q^2)^{1/2}+\alpha \ln F(q)$ where
\begin{equation}
\label{eq.F}
F(q)=\int\int Dx Dy \, e^{[-V(x)-V(x q + y \sqrt{1-q^2})]},
\end{equation}
and $Dx=e^{-x^2/2}dx/\sqrt{2 \pi}$. The authors show that $q=0$
is always a maximum for $\alpha \leq \alpha_{lb}$. In the case of highly 
anisotropic data distributions, when the learning task 
is simple enough that only the variance matters, this leads to 
$\alpha^* = \alpha_{lb}$. However, they disregarded the possibility of having
other extrema, which we expect to
exist~\cite{GoBu,BuGo} if there is some structure in the data along
$\lambda$. We show here that the global maximum may jump
from $q^*=0$ to a {\it finite} value $q^* \equiv q_1 $, at
$\alpha^*=\alpha_1 < \alpha_{lb}$. Consider data whith components along $\lambda$ 
drawn according to
\begin{equation}
\label{eq.2gauss}
P(\lambda)=\frac{1}{2 \sigma \sqrt{2 \pi}}
[e^{-(\lambda-\rho)^2/2\sigma^2} +
e^{-(\lambda+\rho)^2/2\sigma^2}].
\end{equation}
A straightforward calculation shows that $G_{\alpha}(q)$ has
indeed a maximum at $q_1>0$, which may overcome the one at $q=0$
for some values of $\rho$ and $\sigma$. This {\it first order}
phase transition of the upper bound signals the onset of a phase
where learning is possible at $\alpha_1 < \alpha_{lb}$. On the
Figure we represent $\alpha_1$ and $\alpha_{lb}$ as a function of
$\rho$, for $\sigma=0.5$. It may be seen that $\alpha_1(\rho)$
leaves $\alpha_{lb}(\rho)$ with a discontinuous slope at
$\rho=0.7023(7)$ ($\alpha_1=\alpha_{lb}=15.177(9)$,
$q_1=0.900(5)$) (the inset shows the two maxima 
of $G(q)$), but smoothly at 
$\rho=1.338(1)$ ($q_1=0$, $\alpha_1=\alpha_{lb}=0.96$). 
In the latter case, both the second
and fourth order coefficients of the $q$ expansion of
$G_{\alpha}(q)$ vanish at the transition. The other inset represents
$G_{\alpha}(q)$ for $\rho=1$; the transition occurs
at $\alpha_1=4.477(5) < \alpha_{lb}=16$, at which $q^*$ jumps
from $0$ to $q_1 \sim 0.876(4)$.

In some limiting cases, the first order transition may occur with
a jump of $q^*$ from $0$ directly to $q^*=1$ at $\alpha^*=\alpha_1=1$.
This arises, for example, for $P(\lambda)=\frac{a}{2}
\sum_{\tau=\pm 1} \delta(\lambda - \tau \rho) + \frac{1-a}{2}
\sum_{\tau=\pm 1} \delta(\lambda - \tau \rho')$, when $a=0.2$,
$\rho=0.5$, $\rho'=1.4$. 

One of the main conclusions of ref.~\cite{HO}, based on this
upper bound, is that retarded learning exists whenever
$\overline{\lambda}=0$. This conclusion is not invalidated 
by the present analysis: although there is no simple and general 
expression for $\alpha^*$, it can be shown that 
$0 < \alpha^* \leq \alpha_{lb}$.  

\begin{figure}
\centerline{\epsfig{figure=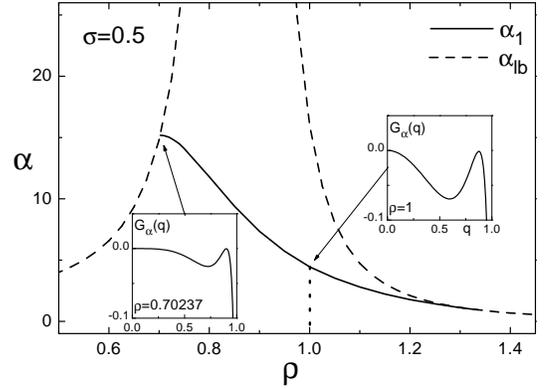,height=5.5cm}}
\caption{\label{fig} Lower bound to the fraction of examples
$\alpha$ below which learning is impossible, as a function of
$\rho$ for $\sigma=0.5$.}
\end{figure}

Acknowledgements. This work was done at the ZiF, Bielefeld,
during the Research Year ``The Sciences of Complexity".

\noindent Arnaud Buhot \\
Theoretical Physics, University of
Oxford, 1 Keble Road, Oxford, OX1 3NP, UK.

\noindent Mirta B. Gordon \\
DRFMC CEA Grenoble, 17 rue des Martyrs,
38054 Grenoble Cedex 9, France.

\noindent Jean-Pierre Nadal \\
LPS ENS\footnote{Laboratory associated with CNRS (UMR 8550) and
the Universities Paris VI and Paris VII}, 24 rue Lhomond, 75231
Paris Cedex 05, France.

\end{document}